\documentclass[3p,two column,fleqn]{elsarticle}
\usepackage{graphicx}
\biboptions{sort&compress}
\usepackage[tbtags]{amsmath}
\usepackage{epsfig}
\usepackage{cprotect,amsfonts,amssymb,bm}
\usepackage{calrsfs}
\usepackage{hyperref}
\journal{Physics Letters B}

\newcommand{\bracket}[2]{\ensuremath{\left\langle#1 \vphantom{#2}\right|\left.\hspace{-3 pt}#2\vphantom{#1}\right\rangle}}

\makeatletter
\newcommand*{\rom}[1]{\expandafter\@slowromancap\romannumeral #1@}
\makeatother

\let\oldhat\hat
\renewcommand{\vec}[1]{\mathbf{#1}}
\renewcommand{\hat}[1]{\oldhat{\mathbf{#1}}}
\DeclareMathAlphabet{\pazocal}{OMS}{zplm}{m}{n}
\begin{document}
	\begin{frontmatter}
	\title{An approach to the quasi-equilibrium state of a self-gravitating system}
	\author[mainaddress]{Azizollah Azizi\corref{correspondingauthor}}
	\ead{azizi@shirazu.ac.ir}
	\author[mainaddress]{Amir A. Khodahami}
	\ead{a.khodahami@shirazu.ac.ir}
	\address[mainaddress]{Department of Physics, Shiraz University, Shiraz 71949-84795, Iran}
	\cortext[correspondingauthor]{Corresponding author}
	\date{Received: date / Accepted: date}
	\begin{abstract}
	We propose an approach to find out when a self-gravitating system is in a quasi-equilibrium state. This approach is based on a comparison between two quantities identifying behavior of the system: a measure of interactions intensity and the area. Gravitational scattering cross section of the system, defined by using the two-particle scattering cross section formula, is considered as the measure of interactions intensity here. A quasi-equilibrium state of such system is considered as a state when there is a balance between these two quantities. As a result, we obtain an equation which relates density and temperature for such a system in the non-relativistic classical limit. This equation is consistent with the TOV equation as expected.
	\end{abstract}
	\end{frontmatter}
\section{Introduction}\label{sec_intro}
It has always been important to know under which conditions a many body system is in its equilibrium state. Maximum entropy principle is expected to give a systematic approach to find out these conditions. Although this is the case in systems with short-range interactions, this approach is faced with essential problems when used in systems with long-range interactions, such as self-gravitating systems. Non-additivity or non-extensivity, inequivalence of ensembles, negative specific heat below a minimum energy which indicates that there exists no stable equilibria (Antonov instability and gravothermal catastrophe), and losing the ergodicity are some of the issues appearing in self-gravitating systems \cite{Taruya2005,Padmanabhan1989,Chavanis2003,Antonov1987,Latora1999,Lynden-Bell1967,Montemurro2003,Ramirez-Hernandez2008,He2010}. Furthermore, it can be shown that the usual thermodynamic equilibrium state will never be attained by self-gravitating systems in the thermodynamic limit, where the number of particles becomes infinitely large. Such systems will attain another different state instead, which in general is non-thermalized and called quasi-equilibrium state. This is just because of the long-range nature of the gravitational interaction \cite{Ourabah2020,Labini2020}.
\par In order to deal with non-extensivity in systems with long-range interactions, the so-called generalized entropies, such as the Tsallis entropy, are suggested. They are generalizations of the well-known Boltzmann--Gibbs entropy and contain it as a special limit \cite{Amigo2018,Balogh2020}. These entropies are not extensive in general and so it seems good to make use of them in such systems; but there are reasons which prevent us to do so. In fact, using a non-extensive entropy by its own definition is meaningless. The Boltzmann-Gibbs form of the entropy uniquely results in a non-extensive entropy in systems with long-range interactions. Using the Boltzmann-Gibbs form of the entropy means that we supposed all the microstates compatible with the given constraints have equal probability to occur. This is a quite reasonable assumption and using any other definition for the entropy leads to non-equiprobability of microstates, which needs an explanation based on some constraints not considered explicitly \cite{Filho2005}. Moreover, the probability distributions obtained from maximization of the Boltzmann-Gibbs form of entropy satisfy the multiplication rule for independent events. This is a basic consistency requirement which in general can be violated by other forms of entropy, such as the Tsallis entropy \cite{Presse2013}. Nevertheless, these types of entropy are useful to understand the way in which some out of equilibrium complex systems behave. Such systems have relatively long lifetimes in the so called meta-equilibrium states, which may be described by generalized entropies \cite{Abe2003,Ehlers2009}.
\par 	Einstein equations of gravitation lead to the hydrostatic equilibrium condition for self-gravitating systems in the stationary limit. So it seems that there is a deep connection between general relativity theory and thermodynamics (see e.g.~\cite{Adami2021}, \cite{Padmanabhan2014} and \cite{Padmanabhan2010}). It can be shown that the results obtained from maximum entropy principle of such systems coincide with those obtained from general relativity equations. For special case of the self-gravitating perfect fluid, one can show that both approaches lead to the Tolman--Oppenheimer--Volkoff (TOV) equation \cite{Gao2011}. Therefore, it seems that there is no need to maximize the entropy of the system when using Einstein equations of gravitation \cite{Sorkin1981,Gao2011,TedJacobson1995,Padmanabhan2004}. However, it should be noted that entropies considered in such works are simply the entropy densities of non-gravitating systems multiplied by the Schwarzschild volume element. So there are non-gravitating particles considered in an external gravitational field, which is obviously a mean-field description. Moreover, there are some issues about stability and causality when combining hydrodynamics and relativity \cite{Bemfica2020,Hoult2021,Noronha2021,Hoult2020}. Although much progress is made in studying self-gravitating systems, there is still no complete understanding of such systems present. Furthermore, whether the statistical mechanics theories and tools can explain the way in which such systems behave, is still a matter of debate. In this situation, some new ideas may shed some light on the problem and help us to find a way which can successfully describe the behavior of self-gravitating systems.
\par Kinetic theory can be used to relate macroscopic and microscopic behavior of thermodynamic systems. Its methods are useful specially when the macroscopic behavior of the system is time-independent. In that case the system is either in an equilibrium state (when only short-range interactions are present) or in a quasi-equilibrium state (when long-range interactions are present). Boltzmann equation is the principal equation governing distribution function for positions and momenta of the particles. It involves three terms sum up to zero when the macroscopic behavior of the system is time-independent: a term due to an external force exerted on particles, another one represents diffusion of particles and last one standing for particles short-range interactions named collision term. This equation describes systems with short-range interactions as well, however, it has difficulties when used in systems with long-range interactions. One can use Vlasov equation in such systems instead. The Vlasov equation is similar to the Boltzmann equation except that the short-range collision term is substituted by a mean field potential term for long-range interactions. The reason is that short-range interactions are suppressed when there are long-range interactions present in the system at thermodynamic limit (where $N\rightarrow\infty$). Reviews and interesting discussions about kinetic theory in systems both with and without long-range interactions can be found in \cite{CREMASCHINI2013,Spohn1991,Sinitsyn2011}. There are useful discussions about entropy production in such systems, governing under the Vlasov equation, and the system's violent relaxation presented in \cite{Farias2018}. The term ``collision'' has two completely distinct meanings in this literature and should be clarified. In kinetic theory, collision is mostly used to indicate a short-range interaction, which can also be indicated by contacting or touching (e.g.~when two solid balls reach each other), and as noted is irrelevant when there are long-range interactions present. At the other hand, particle physicists use the term collision whenever particles interact by exchanging gauge bosons. The Vlasov equation is also called as collisionless Boltzmann equation (collisionless in the sense of kinetic theory) according to the absence of the collision term.
\par In this work, we consider a system of gravitating particles and present an approach to find out when such a self-gravitating system is in a quasi-equilibrium state. This approach is based on collisions among particles and so might result in a fundamental intuition of quasi-equilibrium in such systems. It should be noted that by collision in self-gravitating systems we mean the long-range gravitational collision discussed in the last paragraph. The quasi-equilibrium condition, introduced in Secs.~\ref{eq_cond} and \ref{eq_cond_2}, leads to some restrictions on distribution functions of positions and momenta, which result in a relation between density and temperature in the non-relativistic classical limit discussed in Sec.~\ref{dtr}. Actually the condition restricts the gravitational scattering cross section of the system, which describes the gravitational interactions of particles and is defined in Sec.~\ref{tot_scatt} using the definition of two-particle scattering cross section. Here we only introduce this approach and show that the non-relativistic classical result is acceptable. More precise treatments, including relativistic and quantum considerations, need more jobs to do, which are left for further works on the subject. Nevertheless, at the end of the Sec.~\ref{tot_scatt}, we discuss a bit about relativistic and quantum effects of our approach.
\par There are various textbooks and articles written to describe self-gravitating systems using different approaches and methods---complete discussions are presented in \cite{Binney2008} based on the Newtonian description of gravitation. These approaches are suitable to explain many astrophysical objects. A deeper approach in the subject may need quantum mechanics and relativity---special or even general relativity. Although a quantum version of gravity is not yet released, we use an effective version of the quantum gravity, to enter both quantum and relativity in our approach.
\par Recently, some issues about self-gravitating systems, such as those arise in black hole information problem are discussed in literature (helpful reviews about which are given in \cite{Almheiri2021,Giddings2021} and their references). It seems that present approach can be interpreted in terms of quantum information and its features in gravitational systems. An argument is presented at the end of the Sec.~\ref{eq_cond_2}.
\par This approach may also resolve some problems from the other approaches, which are based on some mean-field descriptions of gravitation. As an example, infinities are arising due to the short behavior of gravitational potential \cite{Robertson2019,Ehlers2009,Follana2000}. In the present approach, using collisions instead of potentials could curb infinities.
\section{Quasi-Equilibrium Condition}\label{eq_cond}
To get into the problem and determine the quasi-equilibrium condition of a self-gravitating system, we consider a particle of it. The particle moves in the system and at the same time interacts with other particles gravitationally. For simplicity, we take particles without any charge or spin and assume that the system is spherically symmetric. So there are only three quantities describing behavior of each particle: its four-momentum, a measure of its interactions (e.g.~its scattering cross section) and volume of the sphere in which it can move. Particles randomly choose their momentum components by a given distribution function. We should consider all the particles if we want to know how the system as a whole behaves. Considering a distribution function for momentum vectors, there remain the other two quantities to identify the particles behavior at each time. It is expected to have a relationship between these two quantities in a quasi-equilibrium state.
\section{Scattering Cross Section of the System}\label{tot_scatt}
In order to describe the collisions among particles, we make use of scattering cross section which is used in particle physics frequently. However, we need to define this quantity for the system as a whole by using scattering cross sections due to the collisions among its constituents. We can simply define it as the sum of scattering cross sections of each pair of particles; i.e.,
\begin{equation}
	\Sigma \equiv \sum_{i < j}^{}\sigma_{ij},
	\label{eq_Sigma}
\end{equation}
where $\sigma_{ij}$ is scattering cross section of particles $i$ and $j$. But as is clear, this definition does not contain correlations and so, is corresponding to addition of individual probabilities without taking care about interferences. In order to take interferences into account, we consider the formula for differential scattering cross section of the pair of particles \cite{Peskin2018}
\begin{equation}
	\begin{split}
	d\sigma_{ij}=&\left(2\pi\right)^{10} \delta^{(4)}\left(p_i+p_j-p^\prime_i-p^\prime_j\right) \frac{\left|\pazocal{M}_{ij}\right|^2}{v_{ij}}\\&\times\frac{1}{\left(2\pi\right)^3}\frac{1}{2E_i}\cdot\frac{1}{\left(2\pi\right)^3}\frac{1}{2E_j}\\&\times\frac{d^3p^\prime_i}{\left(2\pi\right)^3}\frac{1}{2E^\prime_i}\cdot\frac{d^3p^\prime_j}{\left(2\pi\right)^3}\frac{1}{2E^\prime_j}.
	\end{split}
	\label{eq_diff_scatt_cross_sec}
\end{equation}
In the above equation, $p$'s and $p^\prime$'s are particles four-momenta before and  after scattering with their zeroth components specified by $E$ and $E^\prime$ respectively. Also $\pazocal{M}_{ij}$ is scattering amplitude and $v_{ij}$ is magnitude of relative velocity of the particles. We should somehow add these scattering amplitudes before squaring to include interferences in the formula for scattering cross section of the system. But let write the formula for two-particle scattering cross section, Eq.~\eqref{eq_diff_scatt_cross_sec}, in a more appropriate appearance first. Each particle is bounded in a region with volume $V$ and thus, has uncertainty in its momentum vector in the amount of $\delta^3p=(2\pi)^3/V$. Therefore, particles initial momenta are not exactly known and we should explicitly take such uncertainties into account
\begin{equation}
	\begin{split}
	d\sigma_{ij}=&\frac{V^2}{\left(2\pi\right)^6}\cdot\left(2\pi\right)^{10}\delta^{(4)}\left(p_i+p_j-p^\prime_i-p^\prime_j\right)\frac{\left|\pazocal{M}_{ij}\right|^2}{v_{ij}}\\&\times\frac{\delta^3 p_i}{\left(2\pi\right)^3}\frac{1}{2E_i}\cdot\frac{\delta^3 p_j}{\left(2\pi\right)^3}\frac{1}{2E_j}\\&\times\frac{d^3p^\prime_i}{\left(2\pi\right)^3}\frac{1}{2E^\prime_i}\cdot\frac{d^3p^\prime_j}{\left(2\pi\right)^3}\frac{1}{2E^\prime_j}.
	\end{split}
	\label{eq_diff_scatt_cross_sec_uncertainty}
\end{equation}
The above expression is the same as one given in Eq.~\eqref{eq_diff_scatt_cross_sec}, except that it is written in a more symmetric manner with respect to initial and final momenta as a result of entering the momentum uncertainties explicitly. Generalizing Eq.~\eqref{eq_diff_scatt_cross_sec_uncertainty} in the desired way, we propose a formula for differential scattering cross section of a self-gravitating system
\begin{equation}
	\begin{split}
	d\Sigma=&\frac{V^2}{\left(2\pi\right)^6}\cdot\left(2\pi\right)^{10} \delta^{(4)}\left(\sum_n p_n-\sum_n p^\prime_n\right) \\&\times\prod_n\left(\frac{\delta^3p_n}{\left(2\pi\right)^3}\frac{1}{2E_n}\frac{d^3p^\prime_n}{\left(2\pi\right)^3}\frac{1}{2E^\prime_n}\right) \\&\times \left|\sum_{i< j}\frac{\pazocal{M}_{ij\text{,others}}}{\sqrt{v_{ij}}}\right|^2,
		\end{split}
	\label{eq_Proposed_diff_scatt}
\end{equation}
where $\pazocal{M}_{ij \text{,others}}$ is
\begin{equation}
	\pazocal{M}_{ij \text{,others}}=\pazocal{M}_{ij}\prod_{k\neq i,j}\bracket{\vec{p}_k}{\vec{p}^\prime_k}.
	\label{eq_M_ij,others}
\end{equation}
It simply means that we do not involve other particles in the scattering process of particles $i$ and $j$. The Dirac delta function in the proposed formula for differential scattering cross section of the system, Eq.~\eqref{eq_Proposed_diff_scatt}, guarantees the conservation of total energy and momentum of the system. This formula includes cross terms corresponding to the intended interferences.
\par As is clear, we are only considering two-particle scatterings in this work. Three or more particle scatterings are higher orders in perturbation (they are at least $G^2$ times less than two-particle scatterings) and we neglect them. If someone wants to involve them, he/she can use scattering cross section formula for more than two incoming particles and generalize it in the same manner. For instance, we consider only the term due to the scattering process of particles $i$ and $j$
\begin{equation}
	\begin{split}
	d\Sigma^{(2)}=&\frac{V^2}{\left(2\pi\right)^6}\cdot\left(2\pi\right)^{10} \delta^{(4)}\left(\sum_n p_n-\sum_n p^\prime_n\right) \\&\times\prod_n\left(\frac{\delta^3p_n}{\left(2\pi\right)^3}\frac{1}{2E_n}\frac{d^3p^\prime_n}{\left(2\pi\right)^3}\frac{1}{2E^\prime_n}\right) \\&\times \left|\frac{\pazocal{M}_{ij}}{\sqrt{v_{ij}}}\right|^2\prod_{k\neq i,j}\bracket{\vec{p}_k}{\vec{p}^\prime_k}\bracket{\vec{p}^\prime_k}{\vec{p}_k}.
		\end{split}
	\label{eq_only_ij}
\end{equation}
By using the relation
\begin{equation}
	\bracket{\vec{p}}{\vec{q}}=\left(2\pi\right)^3\cdot 2E_p\cdot \delta\left(\vec{p}-\vec{q}\right),
	\label{eq_bra_ket}
\end{equation}
and the fact that $\delta(\vec{p}_k-\vec{p}_k)\delta^3 p_k=1$, due to the momentum uncertainty, we can simplify Eq.~\eqref{eq_only_ij} to obtain Eq.~\eqref{eq_diff_scatt_cross_sec} as expected. It can be checked that in the low density limit, cross terms in Eq.~\eqref{eq_Proposed_diff_scatt} can be ignored (a discussion is presented in \ref{app-Cross}). Thus, correlations disappear in that limit and we arrive at Eq.~\eqref{eq_Sigma} for scattering cross section of the system.
\par
\begin{figure}
	\begin{minipage}{0.38\linewidth}
		\includegraphics[width=\linewidth]{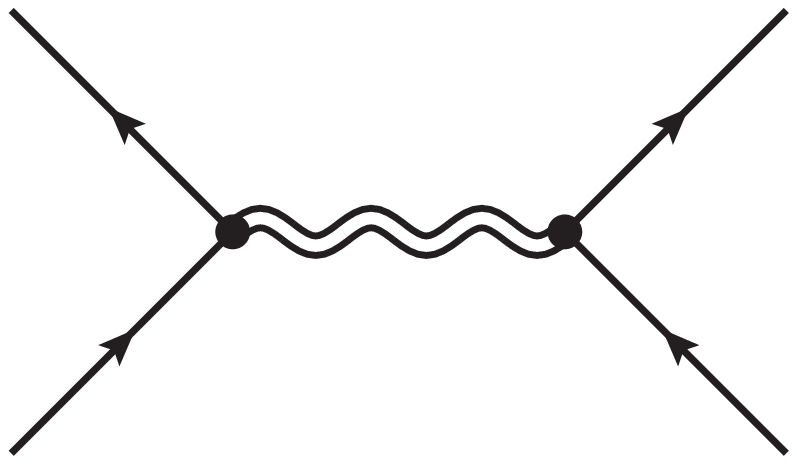}
	\end{minipage}
	\begin{minipage}{0.2\linewidth}
		\begin{center}
		+
		\end{center}
	\end{minipage}
	\begin{minipage}{0.38\linewidth}
		\includegraphics[width=\linewidth]{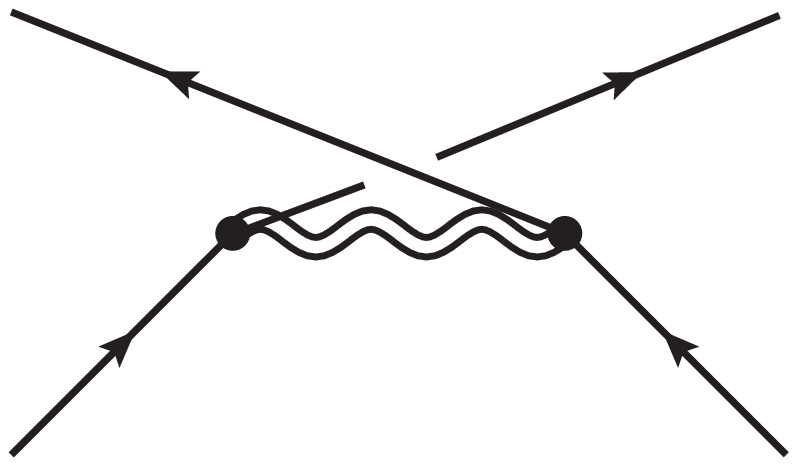}
	\end{minipage}
	\caption{\label{fig1} Tree-level Feynman diagrams contributing to the gravitational scattering process of two indistinguishable particles.}
\end{figure}

It is a customary difficulty to face with infinity when using scattering cross section for a long-range interaction (e.g.~divergence of the total scattering cross section in Rutherford scattering). Such an infinity arises from forward scattering ($\theta\approx0$) and we should consider a cut-off ($\theta_0$) for the polar angle ($\theta$) in order to avoid it. In the present case, at least the radius ($R$) of the system is a natural cut-off for the polar angle, i.e., $\theta_0 = a/R$, where $a$ is a microscopic length in the system. However, we do not specify the cut-off here and it will be canceled in our calculations. Obviously in the limit ($\theta_0\rightarrow0$) the infinity will be recovered. In fact, by considering the cut-off ($\theta_0$), we are putting the difficulties arisen from the long-range nature of gravity aside. It should be noted that we did not specified the theory which is expected to consistently describe gravitation in the quantum limit. This is a subject under research, not yet established. In order to perform calculations here, we will use quantum gravity as an effective field theory description of gravitation (see \cite{Donoghue1995}, \cite{Holstein2006} and its appendices). Due to the classical limit which we want to work in, we consider Feynman diagrams only at tree-level here; so the difficulties arising from non-renormalizability of quantum gravity will not appear in this work. However, the scattering cross section of the system, given by Eq.~\eqref{eq_Proposed_diff_scatt}, includes relativistic and quantum effects when using a quantum gravity theory. In fact, a quantum gravity theory leads to scattering amplitudes, $\pazocal{M}$'s, which include relativistic and quantum effects. For the present case where we use the effective quantum gravity, one can obtain quantum effects even at tree-level. The degeneracy pressure in fermionic systems originates directly from the Fermi statistics (anti-commutation quantization) here, and completely is a quantum mechanical aspect. There are two Feynman diagrams contributing for each pair of particles at tree-level in a fermionic system, as sketched in Fig.~\ref{fig1}. Only the first one is considered in the classical limit which leads to the $1/r$ potential. The reason is that in the classical limit particles are distinguishable and thus we ignore the second diagram which is standing for indistinguishability of particles at the quantum level (see \cite{Peskin2018}, pp.~121-123). However, the second diagram becomes important at the quantum level (where the thermal wavelength of particles is comparable to their mean distance) and is opposite to the first diagram in sign. This is entirely due to the Fermi statistics and will reduce the gravitational scattering cross section of the system. Thus it is opposing to the gravitational attraction which is completely expected. This relative minus sign in fermionic systems is the origin of the degeneracy pressure.
\section{Quasi-Equilibrium Condition: Continued}\label{eq_cond_2}
As mentioned earlier in Sec.~\ref{eq_cond}, there are two quantities identifying behavior of the system. One is a measure of interactions intensity and the other is volume of the sphere to which particles are bounded. We use scattering cross section of the system, defined by Eq.~\eqref{eq_Proposed_diff_scatt}, as a measure of interactions intensity. This quantity has dimensions of $\mathit{length}^2$ and therefore, we can compare it with area of the sphere. Actually it makes sense to compare these two quantities with each other. Considering two systems and assuming that the first one has larger scattering cross section (gravitation is stronger in the first system), if we want both systems to be in their quasi-equilibrium states, the first one should have larger area. In fact, gravitation orders a system and, conversely, increasing area (or volume) of the system disorders it. So in every self-gravitating system there are two quantities opposing to each other: the gravitational scattering cross section ordering the system and the area disordering it. A balance between these two quantities is needed in order to have the system in a quasi-equilibrium state. So for a spherically symmetric system in a quasi-equilibrium state we suggest
\begin{equation}
	\Sigma \propto A.
	\label{eq_Principal_prop}
\end{equation}
Entering a constant $c$, we can write the quasi-equilibrium equation for the system. The constant $c$ should depend on the cut-off ($\theta_0$) in order to remove the infinity arising from the forward scattering. Therefore,
\begin{equation}
	\Sigma =c(\theta_0) \cdot A.
	\label{eq_Principal_equal}
\end{equation}
We call $c(\theta_0)$ resolution constant. Obviously Eq.~\eqref{eq_Principal_equal} performs a global condition. In order to have the system in a quasi-equilibrium state, it should obey a local condition which guarantees all parts of the system to be in their quasi-equilibrium states, so the whole system does too. Therefore, Eq.~\eqref{eq_Principal_equal} should be held on for every $0\leq r\leq R$, where $R$ is radius of the system; i.e.,
\begin{equation}
	\Sigma(r) =c(\theta_0) \cdot A(r).
	\label{eq_Principal_equal_local}
\end{equation}
In the above equation, $A(r)$ is area of the sphere with radius $r$ and $\Sigma(r)$ is scattering cross section of the same sphere, i.e., only particles in the sphere are included in the summation of the scattering cross section formula, Eq.~\eqref{eq_Proposed_diff_scatt}. Finally, Eq.~\eqref{eq_Principal_equal_local} alongside the expression for scattering cross section of the system, Eq.~\eqref{eq_Proposed_diff_scatt}, is our proposed condition for a spherically symmetric system made up of identical particles to be in a quasi-equilibrium state.
\par The area term at the r.h.s. of the proposed equation, Eq.~\eqref{eq_Principal_equal_local}, is a well-known geometric term contributing to the entropy of gravitating systems (see e.g.~\cite{Almheiri2021} and its references). It might be a relation between this approach and the entropy formula, which is left for further works. If it is so, the presented approach might help us to better understand some unresolved problems about black holes. Moreover, it should be noted that this is not the first time that one scales cross section of a gravitating system with its area. Susskind argued that the absorption cross section of incident massless scalar particles on a black hole is related to its area and thus the Bekenstein formula for the black hole entropy is obtained in such a way \cite{Susskind2021}.
\section{Density-Temperature Relation}\label{dtr}
In this section, we obtain an equation which relates density and temperature for a self-gravitating system made up of chargeless scalars as the first solution of the quasi-equilibrium condition, Eq.~\eqref{eq_Principal_equal_local}. We assume that the system is in the non-relativistic classical limit where the density distribution $\rho(r)$ is not too high, so the cross terms in Eq.~\eqref{eq_Proposed_diff_scatt} are negligible, and the particles have medium energies so that the relativistic effects are not important. At tree-level, only diagrams shown in Fig.~\ref{fig1} contribute. Thus the gravitational scattering amplitude of two indistinguishable particles $i$ and $j$ reads as (see e.g.~appendices of \cite{Holstein2006} for Feynman rules)
\begin{equation}
	\begin{split}
	\pazocal{M}_{ij}=&\frac{\kappa^2}{q^2}\Big(m^4+q^2p_i\cdot p_j -2\left(p_i\cdot p_j\right)^2   \\
	& \qquad -2 p_i \cdot p_j\left(p_j-p_i\right)\cdot q \Big),
	\end{split}
	\label{eq_Mij}
\end{equation}
where $\kappa$ is $\sqrt{32\pi G}$, $m$ is particles mass, $p_i$ and $p_j$ are initial four-momenta of the particles and $q$ is the transferred four-momentum from particle $j$ to particle $i$. In the non-relativistic classical limit, particles are distinguishable and momentum vectors can be neglected compared to the mass, thus we obtain
\begin{equation}
	\pazocal{M}_{ij}=\frac{\kappa^2 m^4}{2\left|\vec{q}\right|^2}.
	\label{eq_Mij_approx}
\end{equation}
$\left|\vec{q}\right|$ in the above expression is restricted by the energy conservation equation, which in the non-relativistic limit is
\begin{equation}
	\frac{\left(\vec{p}_i+\vec{q}\right)^2}{2m}-\frac{\left(\vec{p}_i\right)^2}{2m}=\frac{\left(\vec{p}_j\right)^2}{2m}-\frac{\left(\vec{p}_j-\vec{q}\right)^2}{2m}.
	\label{eq_energy_conservation}
\end{equation}
After some elementary algebras one obtains 
\begin{equation}
	\left|\vec{q}\right|=\left(\vec{p}_j-\vec{p}_i\right)\cdot \hat{q}\,.\label{eq_restriction_q}
\end{equation}
The above equation completely determines the amplitude of $\vec{q}$, therefore, the only degrees of freedom are due to its orientation.  There is no preferred direction in the system according to the spherical symmetry assumption, so we take $\hat{q}=\hat{z}$. Positivity of $\left|\vec{q}\right|$ in the above equation restricts $z$ components of the initial momenta
\begin{equation}
	p_{iz}\leq p_{jz}.\label{leq_momenta}
\end{equation}
Each side of Eq.~\eqref{eq_energy_conservation} is the transferred energy, $q_0$, which by making use of Eq.~\eqref{eq_restriction_q} becomes
\begin{equation}
	q_0=\frac{1}{2m}\left(p_{jz}^2-p_{iz}^2\right).
	\label{eq_restriction_q0}
\end{equation}
By inserting the expression of scattering amplitude, Eq.~\eqref{eq_Mij_approx}, into the formula for scattering cross section, Eq.~\eqref{eq_diff_scatt_cross_sec}, and making use of Eqs. \eqref{eq_restriction_q} and \eqref{eq_restriction_q0}, we can find an expression for the scattering cross section of particle $i$ from particle $j$. One can specify its dependencies as follows
\begin{equation}
	\sigma_{ij}=\sigma\left(\vec{p}_i,\vec{p}_j,\theta_0\right).
	\label{eq_sigma_ij_approx}
\end{equation}
Now, we are ready to write an expression for the scattering cross section of the sphere with radius $r$ in terms of the two-particle scattering cross sections. Assuming that the system obeys an undetermined momentum distribution $f(\vec{p})$, we have
\begin{equation}
	\begin{split}
	\Sigma(r) =&\frac{1}{2}\sideset{}{'} \int   \rho(r_1)\rho(r_2) f(\vec{p}_1) f(\vec{p}_2) d^3 \text{r}_1 d^3 \text{r}_2 d^3 \text{p}_1 d^3 \text{p}_2\\&\times\left( \sigma\left(\vec{p}_1,\vec{p}_2,\theta_0\right) +\sigma\left(\vec{p}_2,\vec{p}_1,\theta_0\right)\right).
\end{split}
	\label{eq_Sigma_distribution_prime}
\end{equation}
In the above expression, dummy indices $i$ and $j$ are substituted by $1$ and $2$. It is clear that we should include both $\sigma_{12}$ and $\sigma_{21}$ in order to involve either cases of $\vec{q}$ transferred from 1 to 2 or from 2 to 1. The prime sign indicates the constraint \eqref{leq_momenta} on momenta. Obviously we can keep only $\sigma_{12}$ (or equivalently $\sigma_{21}$) and put the prime sign aside. Therefore, scattering cross section of the sphere is simply
\begin{equation}
	\begin{split}
	\Sigma(r) =& \frac{1}{2}\int \rho(r_1)\rho(r_2) f(\vec{p}_1) f(\vec{p}_2) d^3 \text{r}_1 d^3 \text{r}_2 d^3 \text{p}_1 d^3 \text{p}_2 \\
	&\qquad \times \sigma\left(\vec{p}_1,\vec{p}_2,\theta_0\right).
	\end{split}
	\label{eq_Sigma_distribution}
\end{equation}
In order to make progress in our calculations, we need to specify the functionality of the momentum distribution $f(\vec{p})$. Given that particles are only gravitationally interacting, which is weak, we can choose the momentum distribution as the Maxwell--Boltzmann distribution with a radius dependent temperature
\begin{equation}
	f(p_i)=\frac{1}{\sqrt{2\pi mk_BT(r)}}\exp\left(-\frac{p_i^2}{2mk_BT(r)}\right).
	\label{eq_MB}
\end{equation}
Such a distribution is spherically symmetric as required. Inserting the above distribution into Eq.~\eqref{eq_Sigma_distribution}, we obtain
\begin{equation}
	\begin{split}
	\Sigma(r)=&\int_{0}^{r}\int_{0}^{r}\frac{\rho(r_1)d^3\text{r}_1}{\left(2\pi mk_BT(r_1)\right)^{3/2}}\cdot\frac{\rho(r_2)d^3\text{r}_2}{\left(2\pi mk_BT(r_2)\right)^{3/2}} \\
	&\qquad\quad \times \frac{G^2m^8}{2}\Gamma\left(\frac{T(r_1)}{T(r_2)},\frac{k_BT(r_1)}{m},\theta_0\right),
		\end{split}
	\label{eq_Sigma_Gamma}
\end{equation}
where $\Gamma$ in the above equation is defined as
\begin{equation}
	\begin{split}
	&\Gamma\left(\frac{T(r_1)}{T(r_2)},\frac{k_BT(r_1)}{m},\theta_0\right)=\int  \frac{\sigma\left(\vec{p}_1,\vec{p}_2,\theta_0\right)}{G^2m^8} d^3p_1 d^3p_2\\&\times\exp\left( - \frac{\left|\vec{p}_{1}\right|^2 }{2mk_BT(r_1)}- \frac{\left|\vec{p}_{2}\right|^2}{2mk_BT(r_2)}\right).
	\end{split}
	\label{eq_Gamma}
\end{equation}
$\Gamma$ is a dimensionless quantity which can depend only on $T(r_1)/T(r_2)$ and $(k_BT(r_1))/m$, and is positive and symmetric with respect to $T(r_1)$ and $T(r_2)$. We can approximately take it constant in the limit where temperature is high enough (non-relativistic classical limit) but its variations are not too large. Thus the scattering cross section of the sphere with radius $r$ becomes
\begin{equation}
	\begin{split}
	\Sigma(r)=&\frac{G^2m^8}{2}\Gamma\left(1,\frac{k_BT}{m},\theta_0\right) \\&\times\left[\int_{0}^{r}\frac{\rho(r_1)d^3\text{r}_1}{\left(2\pi mk_BT(r_1)\right)^{3/2}}\right]^2,
	\end{split}
	\label{eq_Sigma_Simplified}
\end{equation}
where $T$ in the above equation is average of the particles temperatures. So the quasi-equilibrium condition, Eq.~\eqref{eq_Principal_equal_local}, results in
\begin{equation}
	G^2m^8\left[\int_{0}^{r}\frac{\rho(r_1)d^3\text{r}_1}{\left(2\pi mk_BT(r_1)\right)^{3/2}}\right]^2 = \frac{8\pi c\left(\theta_0\right)}{\Gamma\left(1,\frac{k_BT}{m},\theta_0\right)}r^2.
	\label{eq_Sigma_A}
\end{equation}
The constant $c(\theta_0)$ completely removes the cut-off from our calculations, so the ratio $(8\pi c(\theta_0))/\Gamma(1,k_BT/m,\theta_0)$ is independent from the cut-off ($\theta_0$) and positive, which we call $D^2(k_BT/m)$. One can use the explicit form of the scattering cross section to obtain that $\Gamma$ scales with $k_BT/m$ in the limit of constant temperature, as is done in the \ref{app-Gamma}. The coefficient is not important due to an arbitrary coefficient in $c(\theta_0)$. Only the ratio of these two coefficients is important, thus we do not need to calculate it. The ratio can be found by using the TOV equation as proceeding, which forces the arbitrary coefficient in $c(\theta_0)$ to be chosen such that our final result agrees with the TOV equation. The above equation holds for all \mbox{$0\leq r\leq R$}, so we take a square root and then a derivative with respect to $r$ to arrive at the following equation between density and temperature
\begin{equation}
	\rho (r)=\frac{D\left(k_BT/m\right)}{Gm^4}\cdot \frac{\left(2\pi mk_BT(r)\right)^{3/2}}{4\pi r^2}.
	\label{eq_dt}
\end{equation}
We can eliminate the coefficient $D(k_BT/m)$ by the assumption that there are $N$ particles in the system, therefore
\begin{equation}
	\rho (r)=\frac{N}{4\pi r^2}\cdot \frac{\left(k_BT(r)\right)^{3/2}}{\int_{0}^{R}\left(k_BT(r^\prime)\right)^{3/2}dr^\prime}.
	\label{eq_dt_notD}
\end{equation}
Comparing Eqs.~\eqref{eq_dt} and \eqref{eq_dt_notD} and making use of the scaling relation $D(k_BT/m)\sim \sqrt{m/k_BT}$, we can see that the quasi-equilibrium condition, Eq.~\eqref{eq_Principal_equal_local}, leads to a balance between kinetic and potential energies in the non-relativistic classical limit, i.e.,
\begin{equation}
	\frac{GM^2}{R}\approx Nk_BT,
	\label{eq_kinetic-potential}
\end{equation}
where $M$ is total mass of the system ($M=Nm$). The obtained balancing of the kinetic and potential energies in the quasi-equilibrium state is completely expected. However, it only makes sense in the classical limit where we make use of potential energies to describe interactions. Notably there is no infinity appearing in this approach due to the short-range behavior of gravity, or equivalently due to an extremely large amount of momentum transferring. The reason is when we use a distribution function for momentum components, we exclude the nonphysical very large momenta.
\par If temperature profile is specified for a system, one can find density from Eq.~\eqref{eq_dt} or Eq.~\eqref{eq_dt_notD}, and pressure by
\begin{equation}
	P(r)=\rho(r)k_BT(r),
	\label{eq_pressure}
\end{equation}
which holds due to the fact that the gravitational interaction is weak. Inserting the obtained relation between density and temperature, Eq.~\eqref{eq_dt}, into the TOV equation in the specified limit (non-relativistic classical limit, small temperature variations and linear gravity), and making use of the Eq.~\eqref{eq_pressure} we see that 
\begin{equation}
	\frac{d}{dr}\left(\rho k_BT\right)=-\frac{Gm^2}{r^2}\rho\int_0^r\frac{D}{Gm^4}\frac{\left(2\pi mk_BT\right)^{3/2}}{4\pi r^{\prime 2}}d^3r^\prime,
	\label{TOV}
\end{equation}
which implies
\begin{equation}
	\frac{-2k_BT}{r^3}=-\frac{Gm^2}{r^4}\frac{D}{Gm^4}\left(2\pi mk_BT\right)^{3/2}r.
	\label{TOV-inserted}
\end{equation}
$r$ is the only variable here due to the spherical symmetry assumption, for which both sides of the above equation behave similarly. Hence the obtained relation between density and temperature, Eq.~\eqref{eq_dt}, agrees with the TOV equation and $D(k_BT/m)$ is obtained as
\begin{equation}
	D\left(\frac{k_BT}{m}\right)=\frac{1}{\pi}\sqrt{\frac{m}{2\pi k_BT}}.
	\label{eq_D}
\end{equation}
It is worthwhile to note that the obtained relation is completely in agreement with our discussions next to the Eq.~\eqref{eq_Sigma_A}. Although we ignored the cross terms of Eq.~\eqref{eq_Proposed_diff_scatt} in deriving the density-temperature relation, Eq.~\eqref{eq_dt}, the discussion presented in \ref{app-Cross} shows how the cross terms alter the relation.
\section{Conclusions}\label{conc}
In this work, we introduced a new approach to study quasi-equilibrium state of a self-gravitating system. This approach is based on a comparison between two quantities which we expect to identify behavior of the system: gravitational scattering cross section and area. We used an expression for two-particle scattering cross section and generalized it in a specific way to obtain a formula for scattering cross section of the system. In the non-relativistic classical limit, a result is obtained in agreement with the TOV equation. This approach is both of conceptual and computational importance. In fact, using particle physics concepts to describe quasi-equilibrium states leads to a fundamental description of such states. At the other hand, this approach includes relativistic and quantum effects due to the quantum field theory considerations. Here we only introduced this approach and showed that its non-relativistic classical result is acceptable. Although the origination of the degeneracy pressure is addressed, deriving it and obtaining other quantum/relativistic effects are postponed to further works. It should also be mentioned that most of the real world self-gravitating systems have a non-zero angular momentum which is not considered here. Actually we considered only the systems with spherical symmetry in this work.
\section*{Acknowledgments}\label{ack}
We thank Mohammad H. Zarei for his helpful tip.
\appendix
\section{Discussions About the Cross Terms In Eq.~\eqref{eq_Proposed_diff_scatt}}\label{app-Cross}
Terms appearing in Eq.~\eqref{eq_Proposed_diff_scatt}, can be divided into three types. Type-0, ones like $\pazocal{M}_{12}\pazocal{M}_{12}$ which were used before in this paper to attain Eq.~\ref{eq_dt_notD} . Type-I, ones like $\pazocal{M}_{12}\pazocal{M}_{23}$, which is common in one index; and type-II, ones like $\pazocal{M}_{12}\pazocal{M}_{34}$, with no common indices. We make use of the symbols $\textcircled{0},\textcircled{1}$ and $\textcircled{2}$ for diagonal terms of type-0, cross terms of type-I and cross terms of type-II respectively. One can find approximate expressions for the ratios $\textcircled{1}/\textcircled{0}$ and $\textcircled{2}/\textcircled{0}$ by using some simple arguments.
\par For the ratio $\textcircled{1}/\textcircled{0}$, we first note that there are $N\choose 2$ terms in $\textcircled{0}$ and $N\choose 3$ terms in $\textcircled{1}$. Thus we have
\begin{equation}\label{eq_app-Cross_ratio10}
	\frac{\textcircled{1}}{\textcircled{0}}\approx\xi\frac{N}{3} \frac{\;\ldots\pazocal{M}_{12\text{,others}}\,\pazocal{M}_{23\text{,others}}}{\ldots\pazocal{M}_{12\text{,others}}^2},
\end{equation}
where $\xi$ is a numerical constant and ``$\ldots$'' stands for not-written factors. By using Eq.~\eqref{eq_M_ij,others} and assuming $\pazocal{M}_{12}\approx\pazocal{M}_{23}\equiv\pazocal{M}$, one obtains
\begin{equation}\label{eq_app-Cross_ratio10_inserted_M}
	\frac{\textcircled{1}}{\textcircled{0}}\approx\xi\frac{N}{3}\cdot\frac{\int d^3p_3^\prime d^3p_1^\prime\bracket{\vec{p}_3^\prime}{\vec{p}_3}\bracket{\vec{p}_1^\prime}{\vec{p}_1}\delta^3p_1\delta^3p_3\pazocal{M}\ldots}{\int d^3p_3^\prime d^3p_1^\prime\bracket{\vec{p}_3^\prime}{\vec{p}_3}\bracket{\vec{p}_3}{\vec{p}_3^\prime}\delta^3p_1\delta^3p_3\pazocal{M}\ldots}.
\end{equation}
According to Eq.~\eqref{eq_bra_ket} and its following discussion, one can see that the uncertainties $\delta^3p_1$ and $\delta^3p_3$ of the numerator will be survived, but the integrations will be canceled. Moreover, the integration over $d^3p_1$ and the uncertainty $\delta^3p_1$ of the denominator will be survived, but the integration over $d^3p_3$ and the uncertainty $\delta^3p_3$ will be canceled. Thus we end up with
  \begin{equation}\label{eq_app-Cross_ratio10_inserted_M_simplified}
  	\frac{\textcircled{1}}{\textcircled{0}}\approx\xi\frac{N}{3}\cdot\frac{\delta^3p_3\ldots}{\int d^3p_1^\prime\ldots},
  \end{equation}
which by making use of $\delta^3p=(2\pi)^3/V$ and the Maxwell-Boltzmann distribution Eq.~\eqref{eq_MB}, becomes
 \begin{equation}\label{eq_app-Cross_ratio10_MB}
	\frac{\textcircled{1}}{\textcircled{0}}\approx\xi\cdot\frac{N}{V}\cdot\left(mk_BT\right)^{-3/2}.
\end{equation}
We absorbed all the numerical constants in $\xi$ in the above expression. By similar arguments, one can obtain
 \begin{equation}\label{eq_app-Cross_ratio20}
	\frac{\textcircled{2}}{\textcircled{0}}\approx\chi\cdot\left(\frac{N}{V}\cdot\left(mk_BT\right)^{-3/2}\right)^2
\end{equation}
where $\chi$ is a numerical constant again. The obtained expressions can be interpreted in the following way: If the thermal de Broglie wavelength of particles, $\lambda_{\mathrm{th}} =\sqrt{2\pi/ mk_BT}$, is much smaller than their mean distance, then the cross terms are unimportant and the classical result is enough. Otherwise, the particles feel each other in a distance less than their de Broglie wavelength, and the quantum effects appear. Therefore, the cross terms should be considered alongside the other quantum effects (e.g.~contribution of the second diagram in Fig.~\ref{fig1}). Hence the cross terms in Eq.~\eqref{eq_Proposed_diff_scatt} alter the density-temperature relation as
\begin{equation}\label{eq_app-dt-alterd}
	\begin{split}
	\rho (r)=&\frac{N}{4\pi r^2}\cdot \frac{\left(k_BT(r)\right)^{3/2}}{\int_{0}^{R}\left(k_BT(r^\prime)
		\right)^{3/2}dr^\prime}\\&\times\left[1+\pazocal{O}\left(\frac{N}{V}\cdot\left(mk_BT\right)^{-3/2}\right)\right].
		\end{split}
\end{equation}
\section{Scaling Relation of $\Gamma(1,k_BT/m,\theta_0)$}\label{app-Gamma}
Here we make use of the explicit form of the scattering cross section to obtain an scaling relation for the function $\Gamma(1,k_BT/m,\theta_0)$ in the non-relativistic classical limit, small temperature variations and linear gravity. First we use the relation between scattering cross section and scattering amplitude, Eq.~\eqref{eq_diff_scatt_cross_sec}, to obtain
\begin{equation}\label{eq_app-Gamma_dsigma}
	\sigma_{12}\sim d\sigma_{12}\sim \frac{1}{p^3}\frac{m}{p^2}\frac{\left|\pazocal{M}_{12}\right|^2}{p/m}\frac{p^6}{m^4},
\end{equation}
where $p$ is a typical scale for momentum and we used the scaling relations $\delta(p)\sim 1/p$, $E\sim m$, $E_1-E_2\sim p^2/m$ and $v\sim p/m$. Using the obtained expression for scattering amplitude, Eq.~\eqref{eq_Mij_approx}, we have
\begin{equation}\label{eq_app-Gamma_M}
	\left|\pazocal{M}_{12}\right|^2\sim \frac{G^2m^8}{p^4}.
\end{equation}
Inserting the above scaling relation into Eq.~\eqref{eq_app-Gamma_dsigma} and simplifying we obtain
\begin{equation}\label{eq_app-Gamma_sigma}
	\sigma_{12}\sim \frac{G^2m^6}{p^4}.
\end{equation}
Now we use Eq.~\eqref{eq_Gamma} to find that
\begin{equation}\label{eq_app-Gamma_Gamma}
	\Gamma\sim \frac{G^2m^6}{p^4}\frac{p^6}{G^2m^8}=\frac{p^2}{m^2}.
\end{equation}
According to the momentum distribution function, Eq.~\eqref{eq_MB}, one can conclude that $p^2\sim mk_BT$, thus finally we obtain
\begin{equation}\label{eq_app-Gamma_gamma_2}
	\Gamma\sim \frac{k_BT}{m},
\end{equation}
or equivalently
\begin{equation}\label{eq_app-Gamma_D}
	D\sim \sqrt{\frac{m}{k_BT}},
\end{equation}
as desired.
%
\end{document}